\documentclass[12pt]{iopart}
\usepackage{graphicx}
\usepackage{url}
\usepackage{cite}

\begin{document}

\title[Bulk heating in RF atmospheric pressure microplasmas]
{Ionization by bulk heating of electrons in capacitive radio frequency atmospheric pressure microplasmas}

\author{T~Hemke$^1$, D~Eremin$^1$, T~Mussenbrock$^1$, 
A~Derzsi$^2$, Z~Donk\'{o}$^2$
K~Dittmann$^3$, J~Meichsner$^3$, 
J~Schulze$^4$}
\address{$^1$Institute for Theoretical Electrical Engineering, 
Ruhr-University Bochum, Germany}
\address{$^2$Institute for Solid State Physics and Optics, Wigner Research Centre for Physics, Hungarian Academy of Sciences, Budapest, Hungary}
\address{$^3$Institute of Physics, University of Greifswald, Germany}
\address{$^4$Institute for Plasma and Atomic Physics, Ruhr-University Bochum, Germany}

\ead{torben.hemke@tet.ruhr-uni-bochum.de}

%%%%%%%%%%%%%%%%%%%%%%%%%%%%%%%%%%%%%%%%%%%%%%%%%
\begin{abstract}
Electron heating and ionization dynamics in capacitively coupled radio 
frequency (RF) atmospheric pressure microplasmas operated in helium are 
investigated by Particle in Cell simulations and semi-analytical modeling. 
A strong heating of electrons and ionization in the plasma bulk due to high 
bulk electric fields are observed at distinct times within the RF period. 
Based on the model the electric field is identified to be a drift field 
caused by a low electrical conductivity due to the high electron-neutral 
collision frequency at atmospheric pressure. Thus, the ionization is mainly 
caused by ohmic heating in this ``$\Omega$-mode". The phase of strongest bulk 
electric field and ionization is affected by the driving voltage amplitude. 
At high amplitudes, the plasma density is high, so that the sheath impedance 
is comparable to the bulk resistance. Thus, voltage and current are about 
45$^\circ$ out of phase and maximum ionization is observed during sheath 
expansion with local maxima at the sheath edges. At low driving voltages, the 
plasma density is low and the discharge becomes more resistive resulting in a 
smaller phase shift of about 4$^\circ$. Thus, maximum ionization occurs later 
within the RF period with a maximum in the discharge center. Significant 
analogies to electronegative low pressure macroscopic discharges operated in 
the Drift-Ambipolar mode are found, where similar mechanisms induced by a 
high electronegativity instead of a high collision frequency have been 
identified. 
\end{abstract}


\maketitle
\pagebreak
%%%%%%%%%%%%%%%%%%%%%%%%%%%%%%%%%%%%%%%%%%%%%%%%%
\section{Introduction}
Microscopic capacitively coupled plasmas operated at atmospheric pressure and 
at a radio frequency of typically 13.56 MHz, e.g. microscopic atmospheric 
pressure plasma jets ($\mu$-APPJ), are frequently used for surface processing 
and medical applications \cite{Schaefer2009,Benedikt2006,Kong2009a,
Iza2008,Weltmann2008}. The non-thermal glow discharge plasma in these sources 
consists of hot electrons and cold heavy particles (ions, neutrals) close to 
room temperature \cite{Knake2008}. Such discharges can be used for sensitive 
surface treatments including human tissue. Atmospheric pressure plasmas avoid 
the necessity of expensive vacuum systems required for processing applications 
in low pressure discharges, while providing a high degree of dissociation and 
an effective generation of reactive species useful for surface treatment 
\cite{Reuter1,Reuter2,Reuter3,Babayan1998,Becker2006,Hemke2011}.

There are different types of APPJs such as coaxial and plane parallel 
configurations \cite{Benedikt2006,Weltmann2008}. The latter design concept is 
based on the plasma jet introduced by Selwyn et al. in 1998 \cite{Selwyn1998} 
and modified by Schulz-von der Gathen et al. \cite{Schulz2007}: The feed gas 
flows between two electrodes separated by a gap of about 1 - 2 mm and driven 
at 13.56 MHz. In the experiment the electrodes are typically made of stainless 
steel and are enclosed by quartz windows including the plasma volume and the 
effluent. In this way direct optical access to the plasma and the effluent is 
provided. Usually the discharge is operated in helium with some optional 
admixture of oxygen and/or nitrogen with typical gas velocities around 
100 m/s \cite{Schaper2009}. 

The generation of reactive species, that determine surface processes in the 
effluent, is caused by electron impact excitation, ionization, and 
dissociation in the plasma volume between the electrodes. Thus, a detailed 
understanding of the dynamics of highly energetic electrons in the plasma is 
essential and provides the basis for any optimization of surface processing 
applications. The electron dynamics in (microscopic) APPJs has been 
investigated experimentally by, e.g. Schulz-von der Gathen et al. 
\cite{Schulz2007,Schaper2009}, Benedikt et al. \cite{Benedikt2010}, and Kong 
et al. \cite{Kong2008} by Phase Resolved Optical Emission Spectroscopy 
\cite{PROES} as well as numerically by Waskoenig et al. \cite{Waskoenig2010} 
and Kong et al. \cite{Kong2006,Kong2009,Iza2007}. Within the RF period several 
maxima of the emission at distinct positions and times have been observed: 
(i) During sheath expansion electrons are accelerated towards the opposing 
electrode and cause excitation/ionization adjacent to the expanding sheath 
edge \cite{Iza2007}. (ii) At the time of maximum sheath voltage excitation and 
ionization by secondary electrons is observed at both sheaths 
\cite{Waskoenig2010,Niermann2011,Niemi2011}. (iii) During sheath collapse 
another excitation maximum is observed in microscopic APPJs at each electrode 
\cite{Kong2008}. This maximum is assumed to be caused by an electric field 
reversal localized at the sheath edge and caused by electron-neutral 
collisions \cite{Schulze2008c}. (iv) Significant excitation and ionization 
inside the bulk at the times of fastest sheath expansion are observed 
\cite{Benedikt2010,Kong2008}. These maxima have been correlated with high bulk 
electric fields, but their exact physical origin is not completely understood. 
At low driving voltage or power, the excitation during sheath expansion is 
typically stronger than the excitation by secondary electrons \cite{Kong2008}. 
Such mode transitions induced by changing the RF voltage amplitude are similar 
to the mode transitions discussed by Belenguer and Boeuf at low pressures 
\cite{Belenguer1990}. 
Here, we investigate the electron heating and ionization dynamics in an 
atmospheric pressure microplasma with plane parallel electrodes driven at 
13.56 MHz, as a function of the RF voltage amplitude, by Particle in Cell 
(PIC) simulations and semi-analytical modeling. The discharge is operated in 
helium. We reveal the origin of the ionization in the bulk and show that the 
ionization maxima adjacent to the collapsing sheaths are not caused by a 
classical localized field reversal under the conditions investigated. Based on 
the analytical model, we demonstrate that the strong ionization in the bulk 
and at the sheath edges is caused by a high electric field inside the bulk at 
the time of maximum current. This high field originates from a low DC 
conductivity due to a high electron-neutral collision frequency at atmospheric 
pressure in the bulk. The phase shift between current and voltage is found to 
be affected by the driving voltage amplitude. Thus, maximum electron heating 
and ionization occur at different times within the RF period depending on the 
voltage amplitude, that also affects the spatial profile of the electron 
heating and ionization rates. We compare our results to low pressure 
macroscopic electronegative capacitive discharges operated in CF$_4$, where 
similar effects are caused by the high electronegativity instead of a high 
collision frequency \cite{Schulze2011a}. We conclude, that a novel mode of 
discharge operation, the $\Omega$-mode, is present in atmospheric pressure 
microplasmas, where ionization is dominated by ohmic heating in the bulk.

The paper is structured as follows: In section~2, the PIC simulation and the 
semi-analytical model to describe the electric field in the bulk are 
introduced. In the third section, the results are presented and discussed. 
Finally, conclusions are drawn in section~4.

\section{Methods}
\subsection{Particle-In-Cell Simulations}
We use a 1d3v (one spatial dimension and three dimensions in velocity space) 
explicit kinetic code based on the particle-in-cell algorithm to study the 
heating and ionization dynamics in capacitive atmospheric pressure 
microplasmas with plane parallel electrodes of identical surface areas 
separated by a gap of 2 mm. Collisions are treated by the Monte-Carlo method. 
The code works in the electrostatic approximation, i.e. $\vec E=-\nabla\phi$. 
The driving voltage waveform is $\tilde{\phi} = \phi_0 \cos(2 \pi f t)$ with 
f = 13.56 MHz and $\phi_0$ = 500 V, 330 V. 
To overcome the limitation of a very small time step -- the electron elastic 
collision frequency has to be resolved -- 
the simulations are accelerated: We use a coarse-sorting algorithm for massive 
parallelization of the code on graphics processing units (GPU) \cite{Mertmann}.
We consider electrons and He$^+$ ions taking cross sections from 
\cite{Lxcat,Phelps} and use an ion-induced secondary electron emission 
coefficient of $\gamma = 0.1$ as well as an electron sticking coefficient at 
the electrodes of 0.5. The gas temperature is set to 350~K.

In order to compare the results obtained for atmospheric pressure microplasmas 
operated in helium to low pressure macroscopic electropositive and 
electronegative discharges we also perform PIC simulations of Ar and CF$_4$ 
discharges with an electrode gap of 1.5 cm operated at 80 Pa. The cross 
sections for Ar are taken from \cite{Phelps}. In case of CF$_4$, we consider 
electrons and the ions CF$_3^{+}$ , CF$_3^-$, F$^-$ using cross sections and 
rate coefficients from \cite{Kurihara2000,Georgieva2003}. A rate 
constant of 5.5$\times 10^{-13} \, \rm m^3 s^{-1}$ is used for 
the ion-ion recombination (CF$_3^+$ + CF$_3^-$, CF$_3^+$ + F$^-$). 
The coefficient for secondary electron emission at the electrodes due to ion 
bombardment is varied in the electropositive case and is set to $\gamma = 0.1$ 
in electronegative CF$_4$. The probability of sticking of electrons at the 
electrodes is assumed to be $0.8$ in these simulations. 
In both the electropositive and the electronegative cases at low pressure 
the gas temperature is kept constant at 350~K \cite{Proshina2010}.

\subsection{Semi-analytical model for the bulk electric field}
In order to understand the physical origin of the bulk electric field obtained 
from the PIC simulations, an analytical expression for the electric field is 
deduced, following \cite{Schulze2008c}.
The model is based on a combination of the electron continuity and momentum 
balance equations, i.e.:
\begin{equation}
\label{eq:e_continuity}
\frac{\partial n_{\rm e}}{\partial t} = 
-\frac{\partial \Gamma_{\rm e}}{\partial x} + S_{\rm e},
\end{equation}

\begin{equation}
\label{eq:e_momentum}
%\frac{\partial (n_{\rm e}v_{\rm e})}{\partial t} = 
%-\frac{\partial n_{\rm e}v_{\rm e}v_{\rm e}}{\partial x} + 
%\frac{1}{m_{\rm e}} \frac{\partial p_{\rm e}}{\partial x} + 
%n_{\rm e} \frac{e}{m_{\rm e}} E = -n_{\rm e} \nu_{c\rm e} v_{\rm e}
\frac{\partial \Gamma_{\rm e}}{\partial t} = 
-\frac{\partial (v_{\rm e} \Gamma_{\rm e})}{\partial x}  
-\frac{1}{m_{\rm e}} \frac{\partial p_{\rm e}}{\partial x} 
-\frac{e}{m_{\rm e}} n_{\rm e} E 
-\nu_{c\rm e} \Gamma_{\rm e}.
\end{equation}

Here, $p_{\rm e}$ is the electron partial pressure, with 
$p_{\rm e} = k_{\rm B} n_{\rm e} T_{\rm e}$, $n_e$ is the electron density, 
$T_{\rm e}$ is the electron temperature, $E$ is the electric field, 
$\Gamma_{\rm e}$ is the electron flux, $m_{\rm e}$ is the electron mass, 
$S_{\rm e}$ is the ionization source, and $\nu_{\rm ce}$ is the frequency 
of elastic electron-neutral collisions. Combining (\ref{eq:e_continuity}) and 
(\ref{eq:e_momentum}) and assuming quasineutrality in the bulk as well as a 
spatially homogeneous electron temperature yields:

\begin{equation}
\label{eq:efield}
E = \frac{m_{\rm e}}{n_{\rm e} e^2} \left(
\frac{\partial j_{\rm e}}{\partial t} 
+ \nu_{c \rm e} j_{\rm e}
+ \frac{1}{e n_{\rm e}^2} \left(j_{\rm e}^2 - j^2_{\rm{th,e}} \right) 
\frac{\partial n_{\rm e}}{\partial x} 
+ \frac{j_{\rm e} S_{\rm e}}{n_{\rm e}} \right).
\label{EqEfield}
\end{equation}

Here, $j_{\rm e} = e n_{\rm e} v_{\rm e}$ is the electron conduction current 
density and $j_{\rm{th,e}} = e n_{\rm e} \sqrt{k_{\rm B} T_e/m_{\rm e}}$. The first and 
third term of equation \ref{EqEfield} correspond to electric fields caused by 
inertia effects, while the second term is a drift field, the fourth term is 
the ambipolar field, and the fifth term corresponds to an electric field 
caused by an ionization source.
%%%%%%%%%%%%%%%%%%%%%%%%%%%%%%%%%%%%%%%%%%%%%%%%%
\section{Results}

\begin{figure}[h!]
\begin{center}
\begin{tabular}{cc}
  \includegraphics[width=0.42\textwidth]{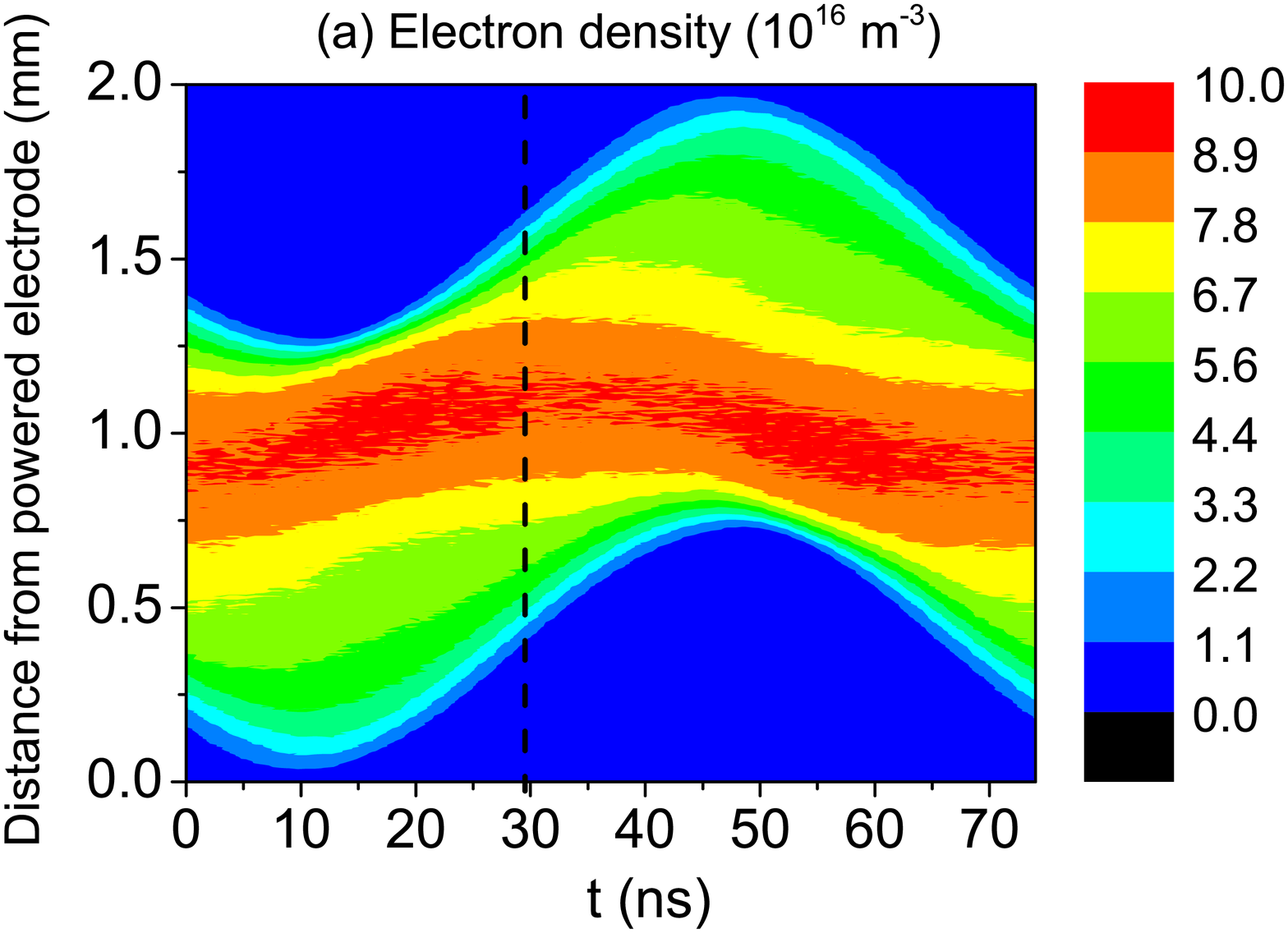}
&
   \includegraphics[width=0.42\textwidth]{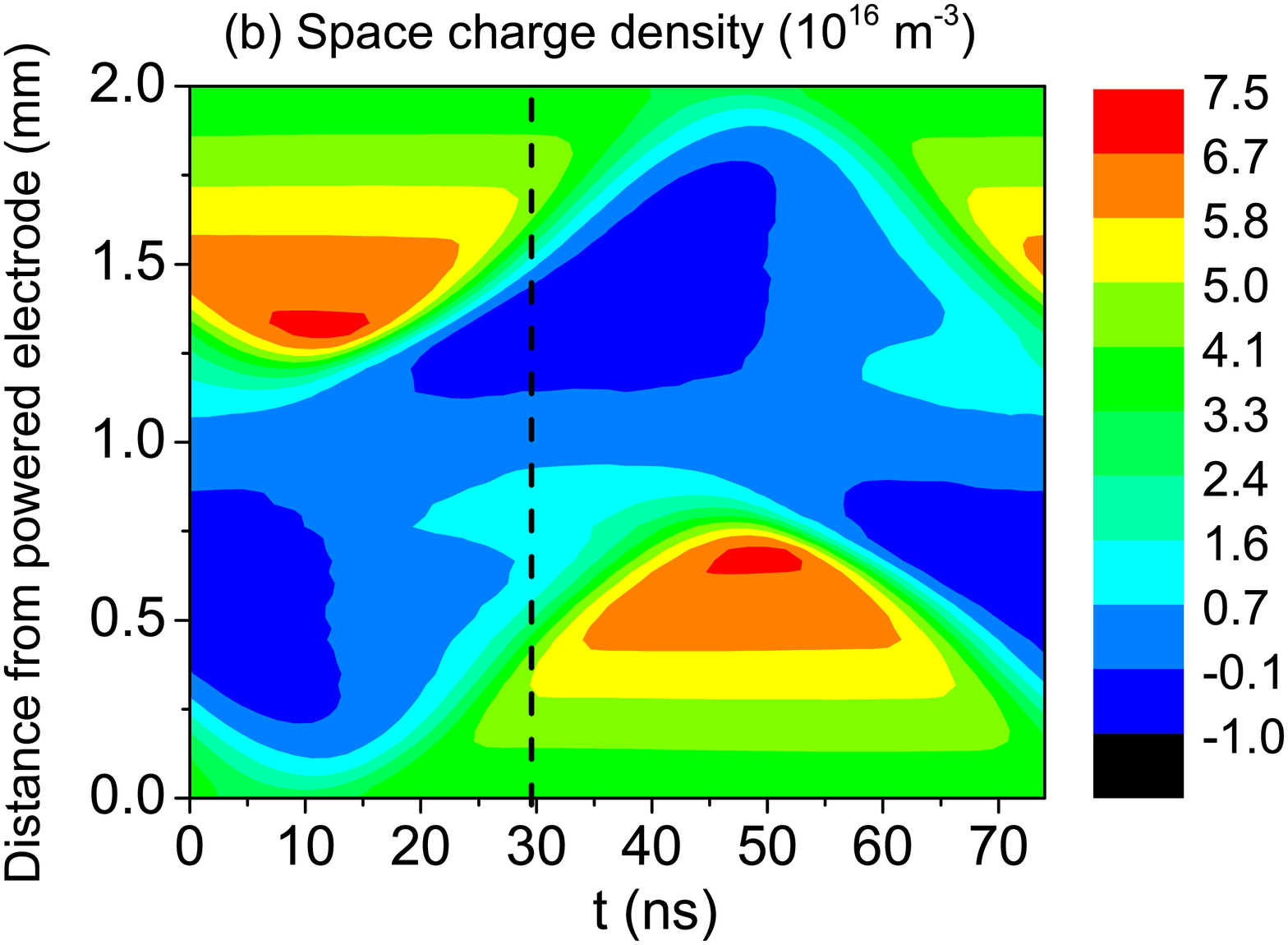}\\
  \includegraphics[width=0.42\textwidth]{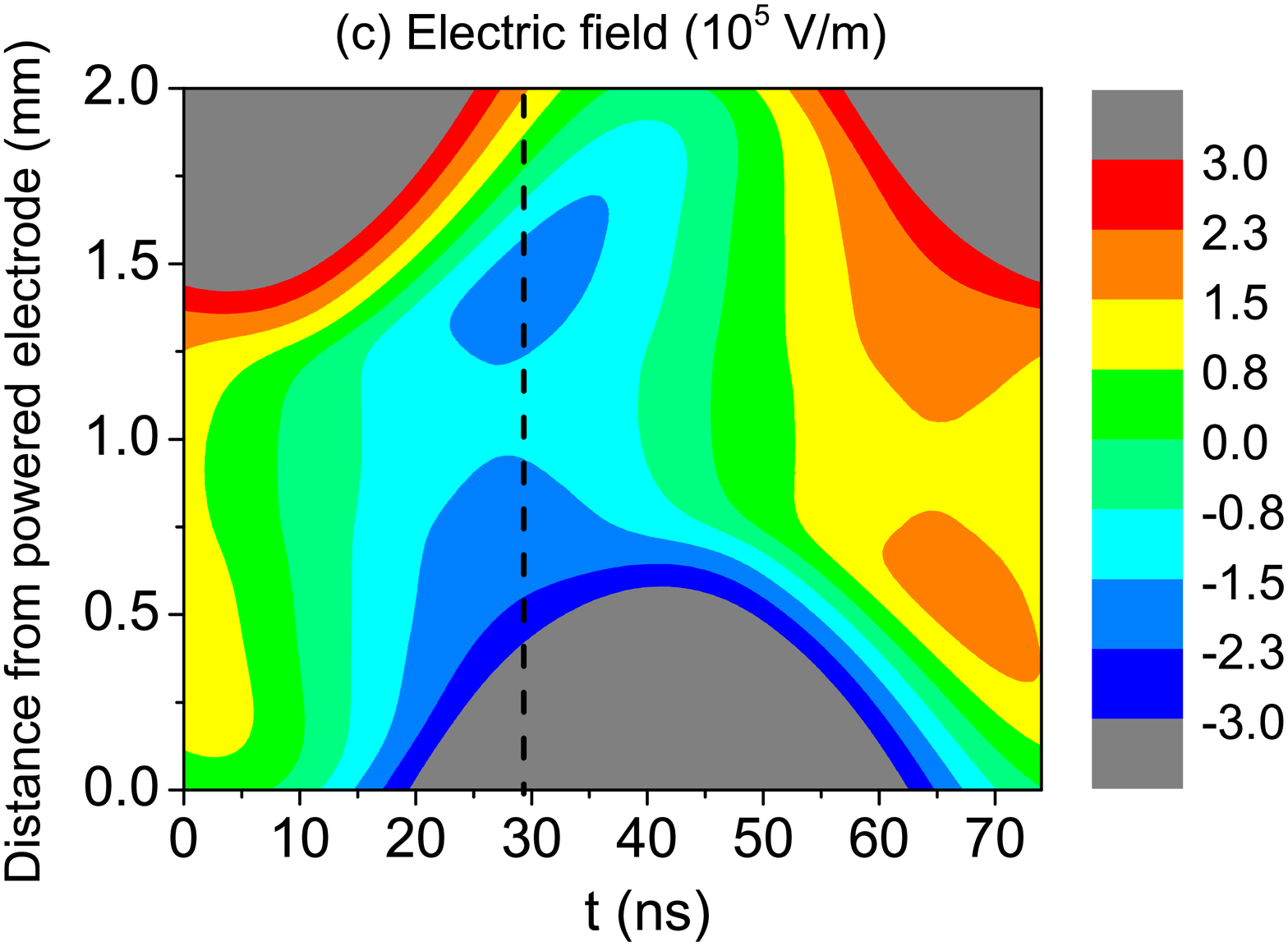}
&
   \includegraphics[width=0.42\textwidth]{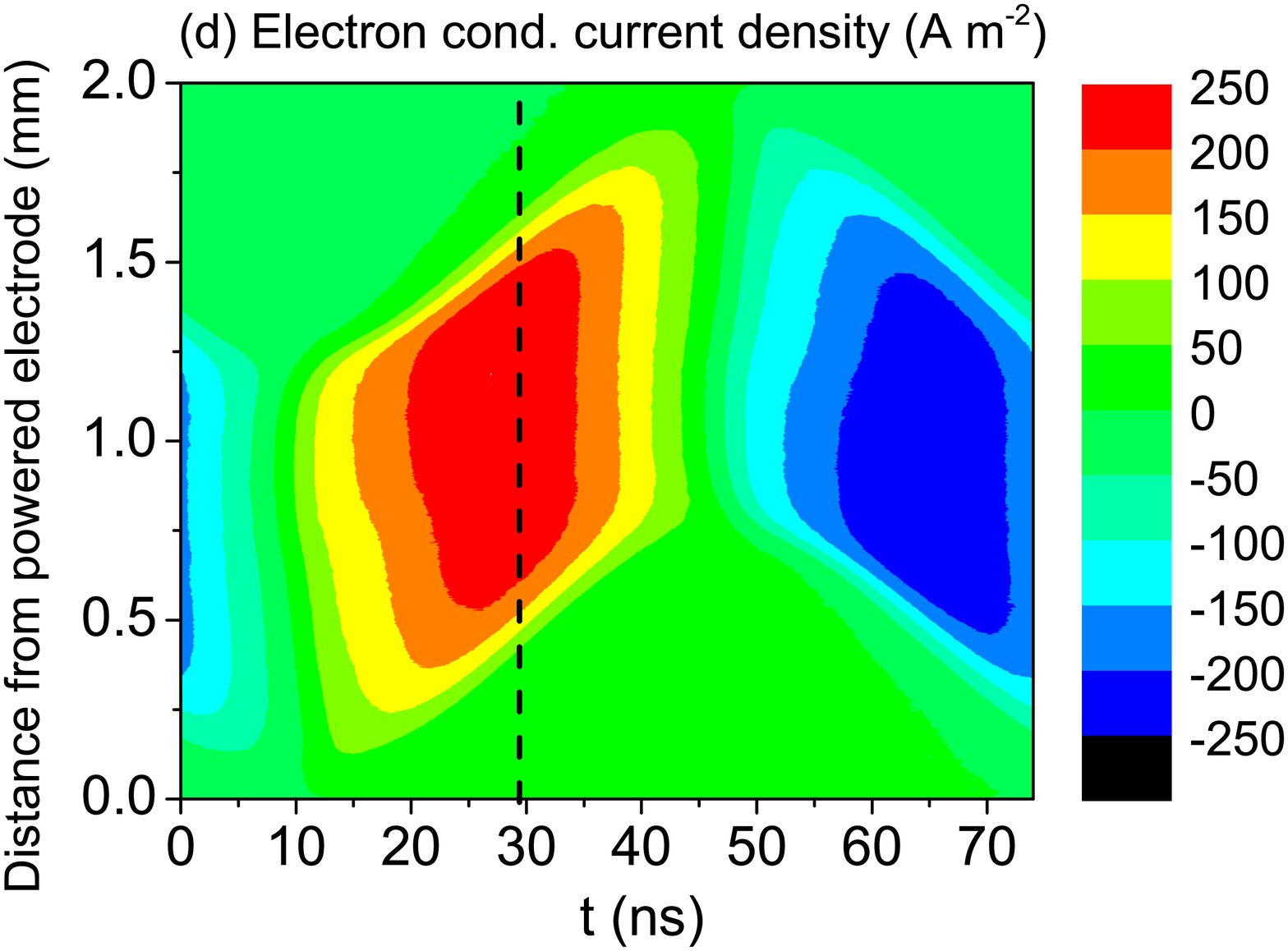}\\
   \includegraphics[width=0.42\textwidth]{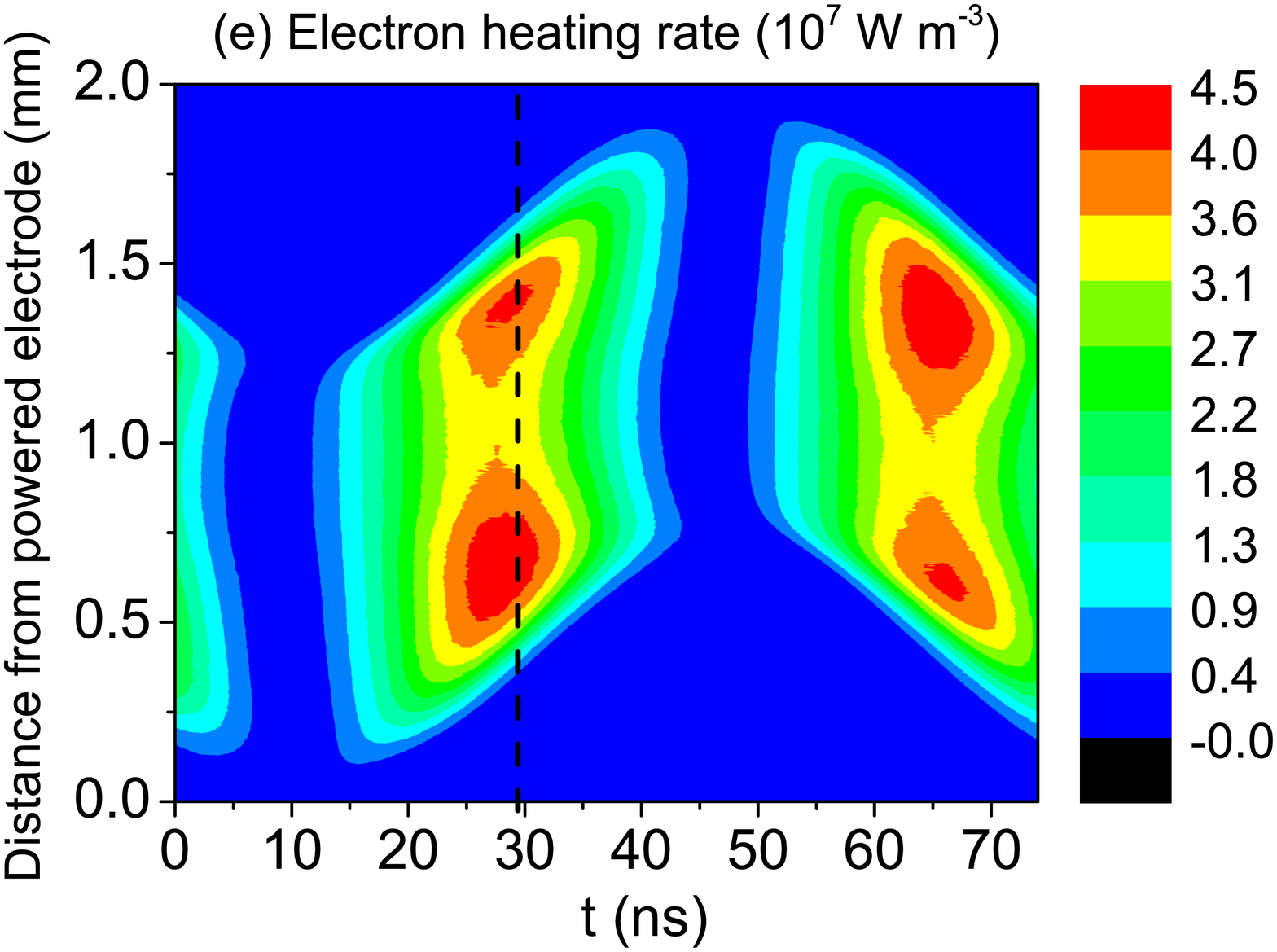}
&
   \includegraphics[width=0.42\textwidth]{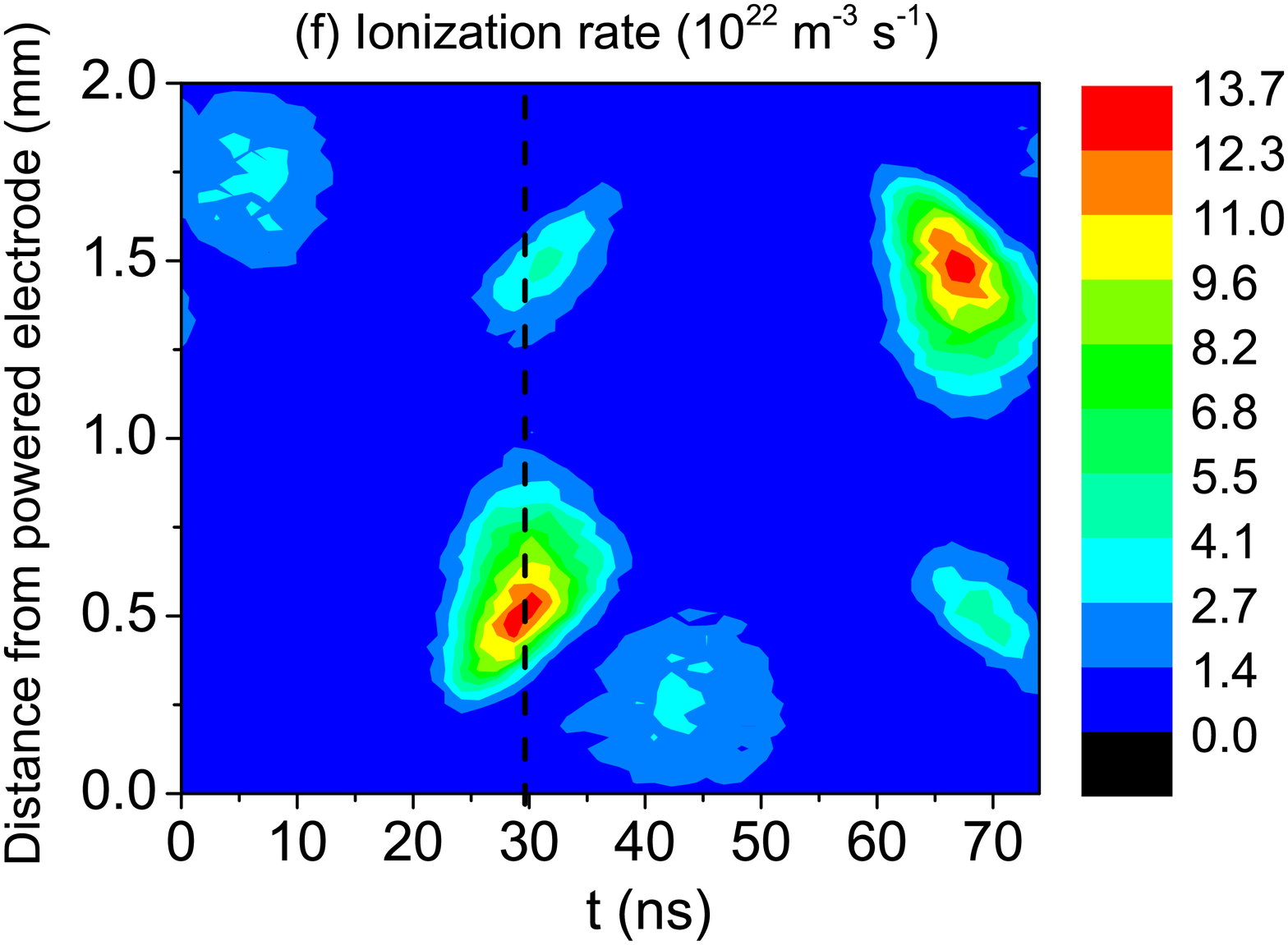}\\
   \includegraphics[width=0.42\textwidth]{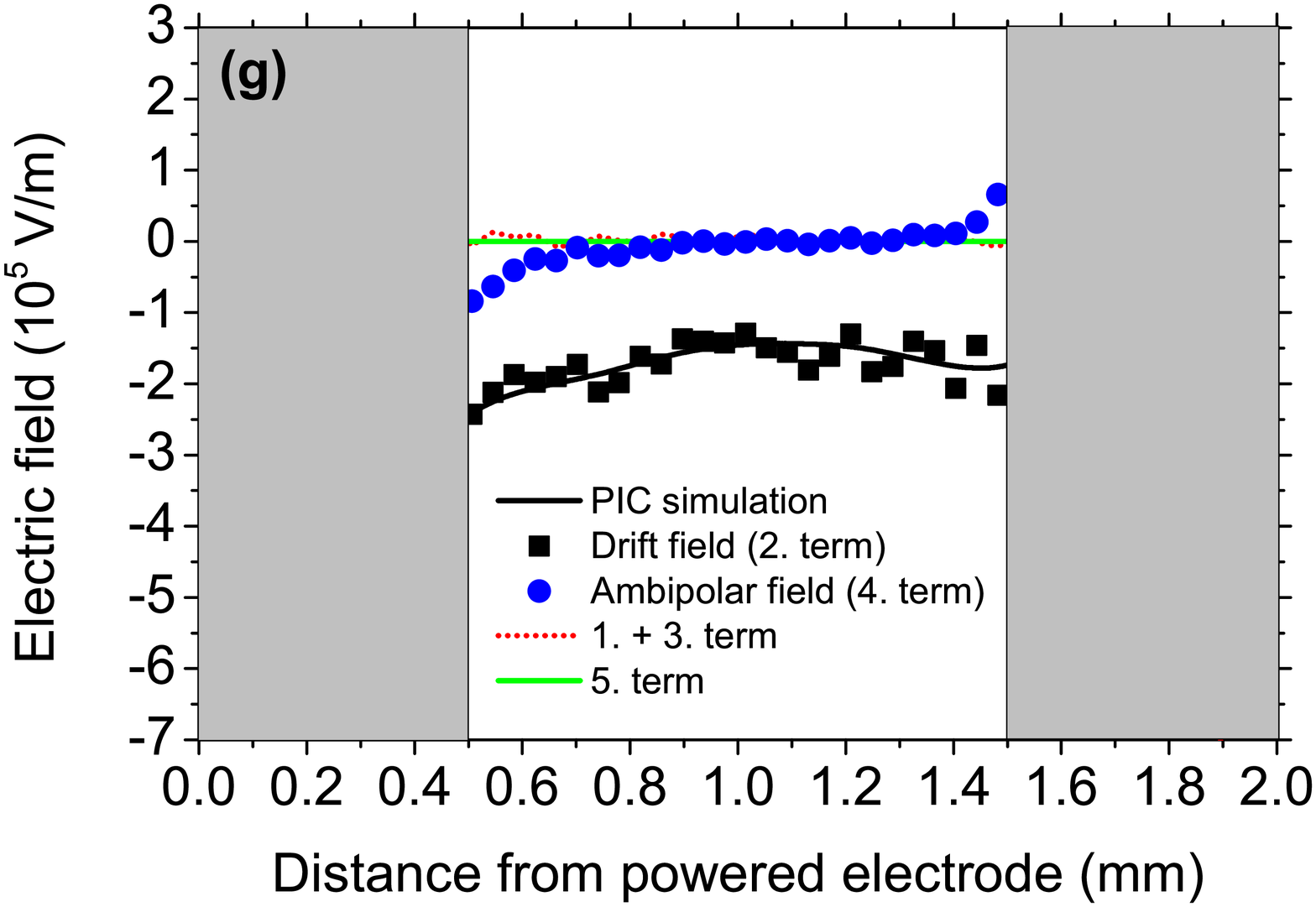}
&
   \includegraphics[width=0.42\textwidth]{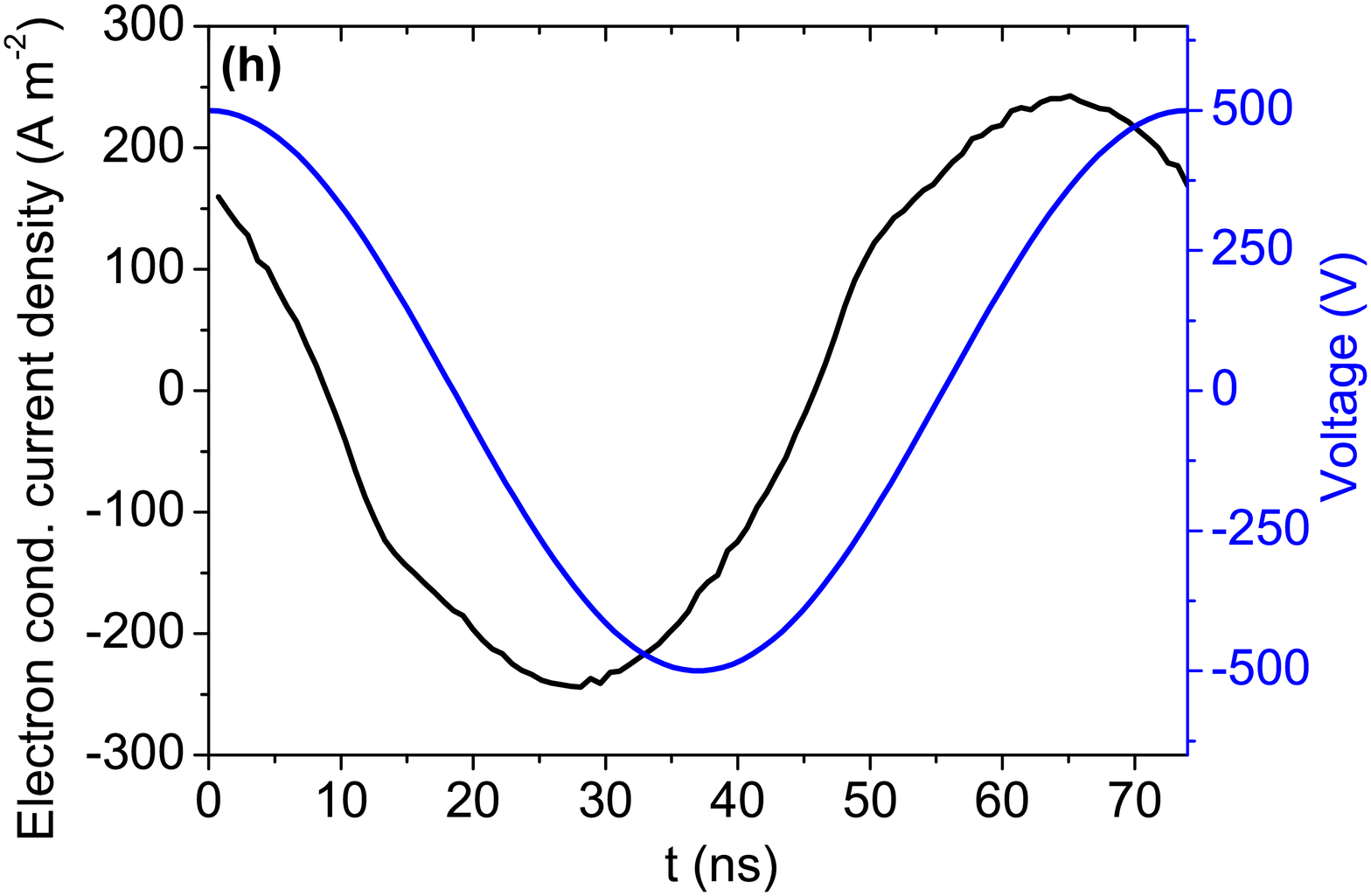}
\end{tabular}
\caption{PIC simulation results (He, 1 atm, 13.56 MHz, 2 mm gap, 500 V): 
Spatio-temporal plots of the (a) electron density, (b) space charge density, 
(c) electric field, (d) electron conduction current density, 
(e) electron heating, and (f) ionization rate. (g) shows the electric field 
profile at the time of max. ionization [vertical dashed lines in (a) - (f)] 
and the individual terms of eq. (\ref{EqEfield}). (h) shows the 
conduction current density in the discharge center and the applied voltage as 
a function of time.}
\label{Plots500V}
\end{center}
\end{figure}

Figure \ref{Plots500V} shows different plasma parameters resulting from PIC 
simulations of capacitively coupled microscopic atmospheric pressure plasmas 
operated in helium with an electrode gap of 2 mm at 13.56 MHz and a voltage 
amplitude of $\phi_0$ = 500 V. The spatio-temporal distributions of the 
electron density, space charge density, electric field, electron conduction 
current density, electron heating rate, and ionization rate obtained from the 
simulation are shown in plots (a) - (f), respectively. The electric field 
profile at the time of maximum ionization marked by vertical dashed lines in 
(a) - (f) resulting from the simulation and the individual terms of 
equation (\ref{EqEfield}) are shown in plot (g). As the model is only valid in 
regions of quasineutrality, its results are not shown inside the sheaths. The 
first and third term of equation (\ref{EqEfield}) are added to limit the 
number of individual lines in the plot. Their sum and each individual value 
are close to zero everywhere in the discharge. The electron conduction current 
density in the discharge center and the applied driving voltage waveform are 
shown in plot (h).

\begin{figure*}[floatfix,h!]
\includegraphics[width=1\textwidth]{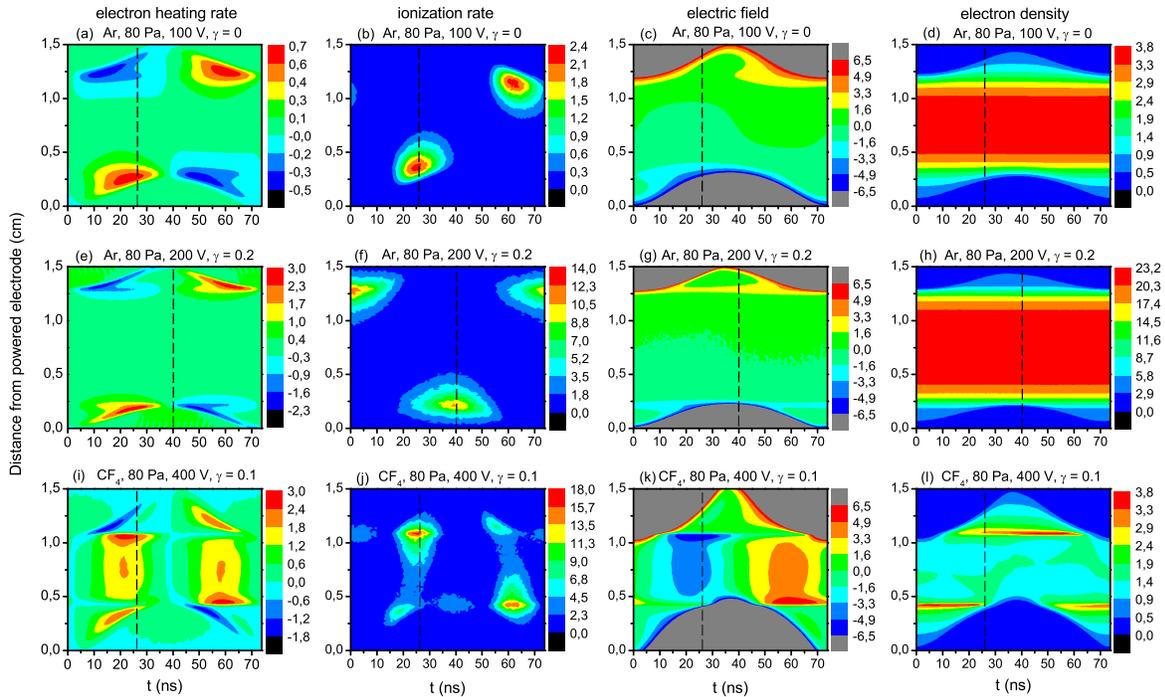}
\caption{\label{fig:1} PIC simulation results: spatio-temporal plots of the 
electron heating rate (first column), ionization rate (second column), 
electric field (third column), and electron density (fourth column) in Ar and 
CF$_4$ discharges driven at 13.56 MHz and 80 Pa with an electrode gap of 
1.5 cm. First row: Ar, 100 V, $\gamma = 0$. Second row: Ar, 200 V, 
$\gamma = 0.2$. Third row: CF$_4$, 400 V, $\gamma = 0.1$. The color scales 
are given in units of $10^5$ W m$^{-3}$ (heating rate), 
$10^{21}$ m$^{-3}$ s$^{-1}$ (ionization rate), $10^3$ V m$^{-1}$ (electric 
field), and $10^{15}$ m$^{-3}$ (electron density).
Reproduced from Ref. \cite{Schulze2011a}.}
\label{HeatModes}
\end{figure*}

\begin{figure}[h]
\begin{center}
\begin{tabular}{cc}
  \includegraphics[width=0.5\textwidth]{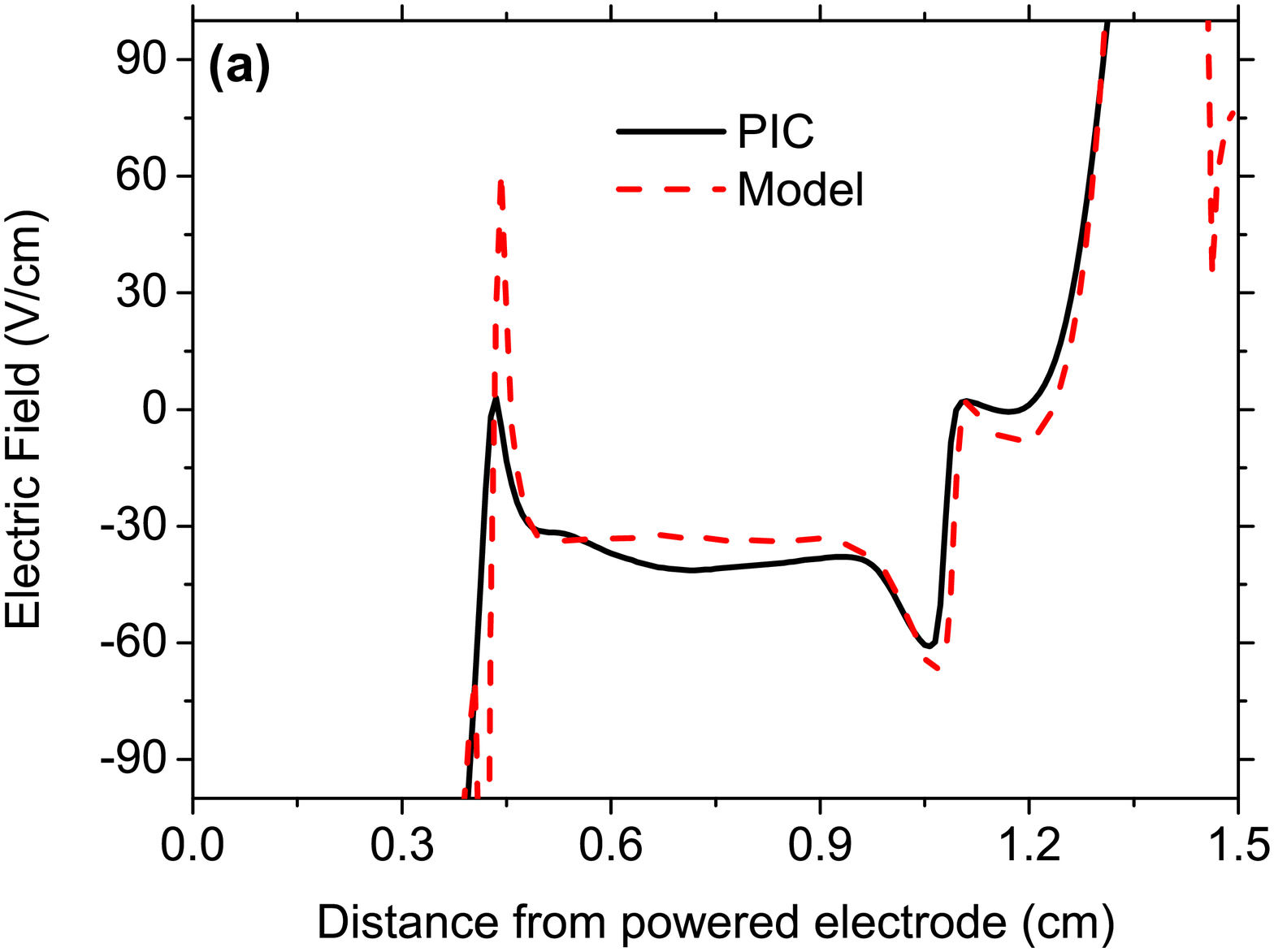}
&
   \includegraphics[width=0.5\textwidth]{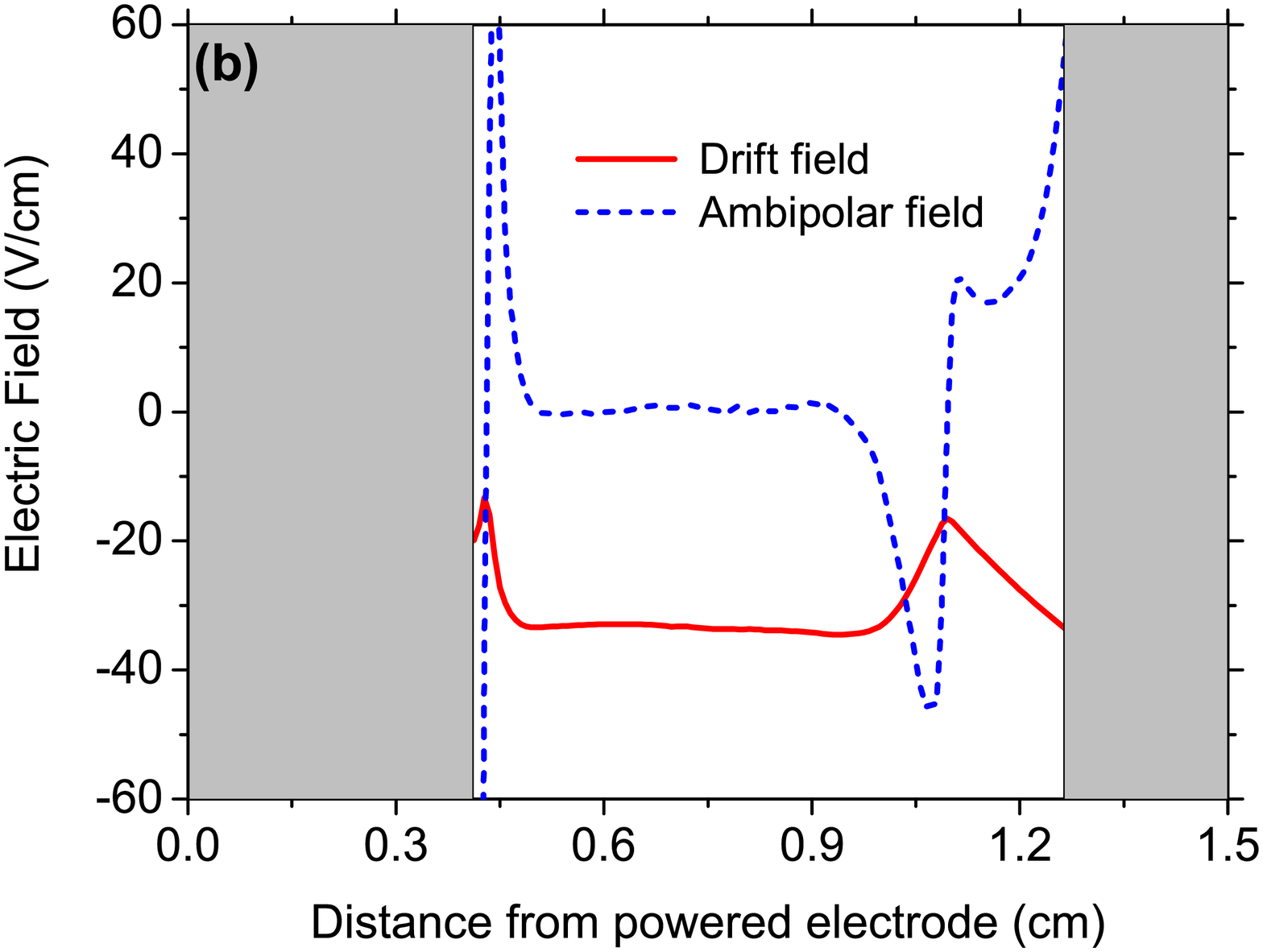}
\end{tabular}
\caption{Profiles of (a) the electric field obtained from the simulation 
(black solid line) and (\ref{EqEfield}) (dashed red line); (b) the second 
(drift field) and fourth (ambipolar field) terms of (\ref{EqEfield}) in the 
bulk at $t \approx 26 \, \rm ns$ [vertical dashed lines in 
Fig.~\ref{HeatModes}(i)-(l)]. Discharge conditions: CF$_4$, 13.56 MHz, 400 V, 
80 Pa, $\gamma = 0.1$. Vertical dashed lines indicate the time of maximum 
ionization. Reproduced from Ref. \cite{Schulze2011a}.} 
\label{ModelCF4}
\end{center}
\end{figure}

Analogous simulation results for the spatio-temporal electron heating rate, 
ionization rate, electric field, and electron density obtained in macroscopic 
low pressure capacitively coupled RF (CCRF) discharges operated in argon and 
CF$_4$ at 13.56 MHz and 80 Pa with an electrode gap of 1.5 cm are shown in 
figure \ref{HeatModes}. The first row shows results for a discharge operated 
in argon at 80 Pa, 100 V voltage amplitude, and $\gamma = 0$, the second row 
corresponds to an argon plasma at 80 Pa, 200 V, and $\gamma = 0.2$, and the 
third row shows results for CF$_4$ at 80 Pa, 400 V, $\gamma = 0.1$. The rows 
of figure \ref{HeatModes} correspond to the 3 different modes of electron 
heating known in macroscopic low pressure capacitive RF plasmas: The first 
row corresponds to the $\alpha$-mode, where electrons are heated by direct 
interaction with the expanding sheaths and cooled by direct interaction with 
the collapsing sheaths \cite{SchulzeStochHeat,SchulzeBeams,Turner}. This 
results in maximum ionization at distinct times within each RF period at 
about 26 ns and 63 ns. The first maximum is indicated by a vertical line in 
plots (a)-(d) in the first row of figure \ref{HeatModes}.
If the secondary yield, $\gamma$, is increased, the discharge is operated in 
$\gamma$-mode at otherwise similar conditions and the ionization is dominated 
by secondary electrons generated by ion impact at the electrodes. These 
$\gamma$-electrons are accelerated and multiplied effectively inside the 
sheaths at the times of maximum sheath voltage, i.e. at about 3 ns and 40 ns 
(vertical dashed line in the second row,\cite{Donko_Gamma,Schulze_Gamma}).

Electronegative discharges can be operated in the Drift-Ambipolar (DA) mode, 
where ionization caused by strong electric fields inside the bulk at distinct 
times within the RF period dominates. This mode is observed in CF$_4$ 
discharges at 80 Pa such as shown in the third row of figure 2 
\cite{Schulze2011a}. 

In microscopic atmospheric pressure plasmas (figure \ref{Plots500V}), the 
electron density peaks in the central bulk region and decreases monotonically 
towards the electrodes similar to electropositive macroscopic CCRF discharges 
(1st and 2nd rows in figure \ref{HeatModes}). In contrast to low pressure 
electropositive discharges the electric field is high inside the plasma bulk 
at two distinct times within the RF period. One phase of high bulk field is 
marked by a vertical dashed line in figure \ref{Plots500V}. The second phase 
of high bulk electric field occurs half an RF period later. At these times, 
the conduction current is high in the discharge center and maximum electron 
heating as well as ionization adjacent to the sheath edges are observed. Weak 
additional ionization by secondary electrons is observed inside the sheaths at 
times of maximum sheath voltage. This is similar to low pressure 
electronegative discharges, where also a high bulk electric field is found at 
distinct phases in the RF period (3rd row in figure \ref{HeatModes}).

Although the spatio-temporal ionization dynamics during sheath expansion in 
atmospheric pressure microdischarges [figure \ref{Plots500V} (f)] looks 
similar to results obtained in low pressure macroscopic CCRF discharges 
operated in $\alpha$-mode [figures \ref{HeatModes} (b)], the physical 
mechanisms causing these maxima are different. At low pressures, electrons 
directly interact with the expanding sheaths and are heated stochastically. 
Electron beams \cite{SchulzeBeams} are generated by the expanding sheaths and 
propagate into the bulk, where they cause ionization typically within a 
distance of one electron mean free path away from the sheath edge. This direct 
interaction of electrons with the time dependent sheath electric field results 
in cooling during sheath collapse. At atmospheric pressure, the electron mean 
free path is below 1 $\mu m$. However, ionization at the time of sheath 
expansion is observed up to 500 $\mu m$ away from the expanding sheath edge. 
This ionization is caused by a high bulk electric field such as shown in 
figure 1 (c), which has local extrema adjacent to the sheath edges at both 
electrodes. The semi-analytical model [equation (\ref{EqEfield})] reproduces 
the electric field well at the time of maximum ionization using input 
parameters from the simulation in terms of $j_{\rm e}$, $n_{\rm e}$, 
$T_{\rm e}$, and $\nu_{\rm ce}$. By separating the contributions of the 
individual terms in equation (\ref{EqEfield}), the physical origin of this 
high bulk field is revealed. It turns out, that the second term (drift field) 
dominates compared to the others. 

Thus, the results of the model and the simulation 
%the comparison between model and simulation 
show [figure \ref{Plots500V} (g)], that the high bulk field is predominantly a 
drift field caused by a low conductivity due to the high electron-neutral 
collision frequency at atmospheric pressure. This field increases slightly 
from the discharge center towards the sheaths due to the decreasing plasma 
density at constant conduction current density. Any slope of the electric 
field profile is correlated with a small local space charge shown in figure 
\ref{Plots500V} (b) \cite{Sakiyama}. Electrons are accelerated in this high 
bulk electric field and cause ionization, where the field is maximum, i.e. at 
the sheath edges. This also explains the local maximum of the ionization at 
the collapsing sheath edge, where no localized field reversal is present. 
Thus, the ohmic bulk heating of electrons in this $\Omega$-mode dominates the 
electron heating and ionization dynamics under these conditions. A remarkable 
difference to the electron heating dynamics in low pressure macroscopic CCRF 
discharges is the absence of electron cooling by the collapsing sheaths. This 
verifies the absence of direct interaction of electrons with the moving 
sheaths and the importance of the bulk heating.

Under the conditions investigated here, current and voltage are 46.8$^\circ$ 
out of phase due to the high plasma density, which causes the sheath impedance 
to be comparable to the bulk resistance, 
i.e. $1/(\omega_{RF} C_s) \approx R_b$. Here
, $\omega_{RF} = 2 \pi \cdot 13.56 \, \rm MHz$, 
$C_s \approx \varepsilon_0 A/\bar{s}$ is the effective capacitance of both 
sheaths determined from the time averaged sheath width, $\bar{s}$, and 
$R_b = (\nu_{ce} d_b)/(\omega^2_{pe} \varepsilon_0 A)$ is the bulk resistance 
determined from the the bulk length, $d_b$, and the electron plasma frequency, 
$\omega_{pe}$ \cite{LLBook}. $A$ is the electrode area.

The presence of a high bulk electric field in electropositive atmospheric 
pressure microdischarges is similar to the situation in low pressure 
macroscopic and electronegative CCRF discharges operated in the 
Drift-Ambipolar mode (3rd row in figure \ref{HeatModes}).
Although these two types of capacitively coupled discharges differ 
significantly in pressure and charged particle species there is a low DC 
conductivity in the bulk, 
$\sigma_{\rm DC} = e^2 n_{\rm e}/m_{\rm e} \nu_{\rm ce}$, in both cases.
However, the origin of the low conductivity is different.
It can be explained by a high elastic collision frequency $\nu_{\rm ce}$ 
in atmospheric pressure CCRF plasmas, while it is caused by a low electron density 
$n_{\rm e}$ due to the high electronegativity in low pressure macrocopic 
electronegative CCRF discharges  \cite{Schulze2011a}. 
In the latter case the second and fourth terms of equation (\ref{EqEfield}) 
reproduce the high bulk electric field such as shown in figure \ref{ModelCF4}. 
The shape of the ambipolar field is caused by the presence of local maxima of 
the electron density in the electropositive edge regions of this 
electronegative plasma [figure \ref{HeatModes} (l)].

\begin{figure}[h!]
\begin{center}
\begin{tabular}{cc}
  \includegraphics[width=0.39\textwidth]{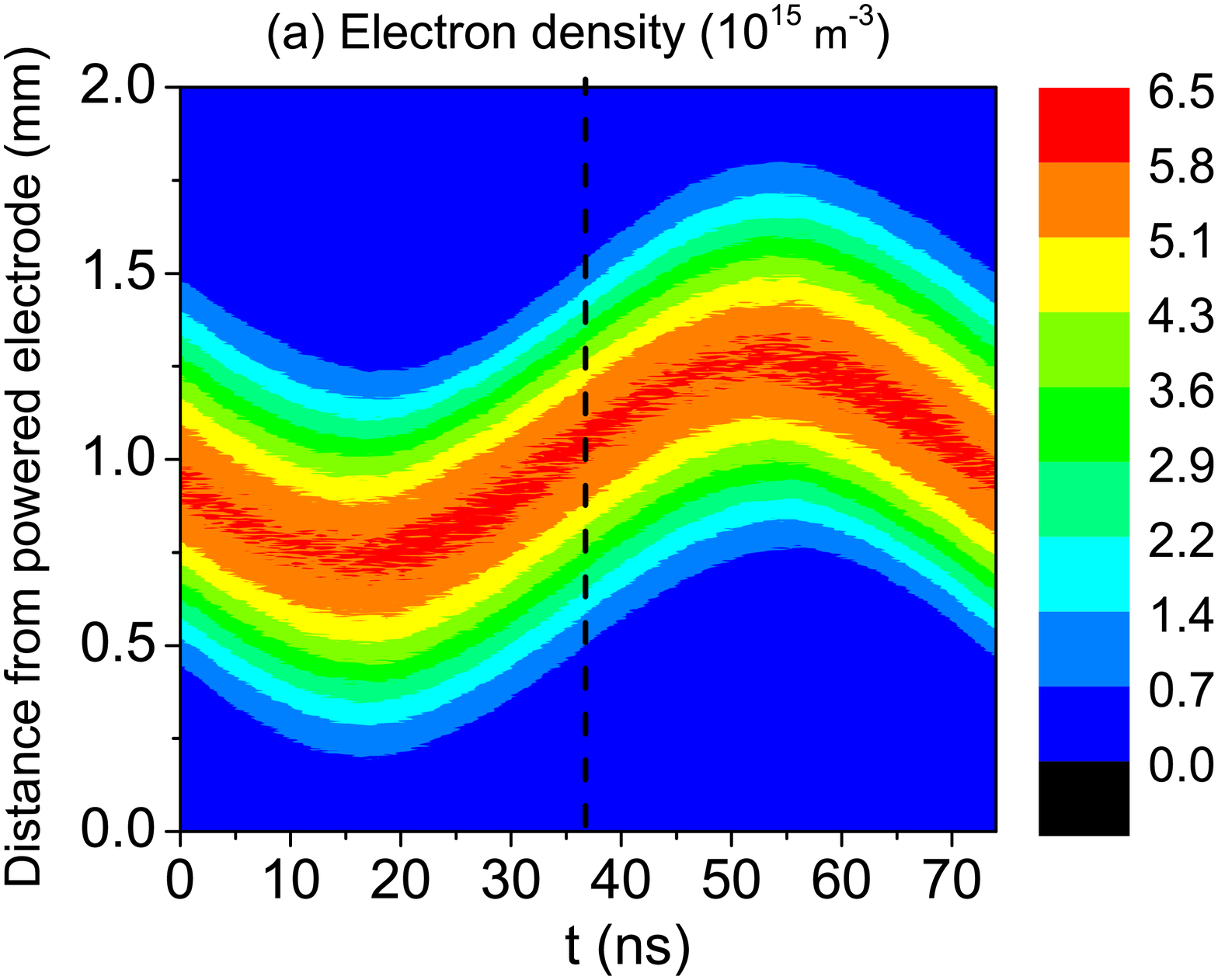}
&
   \includegraphics[width=0.39\textwidth]{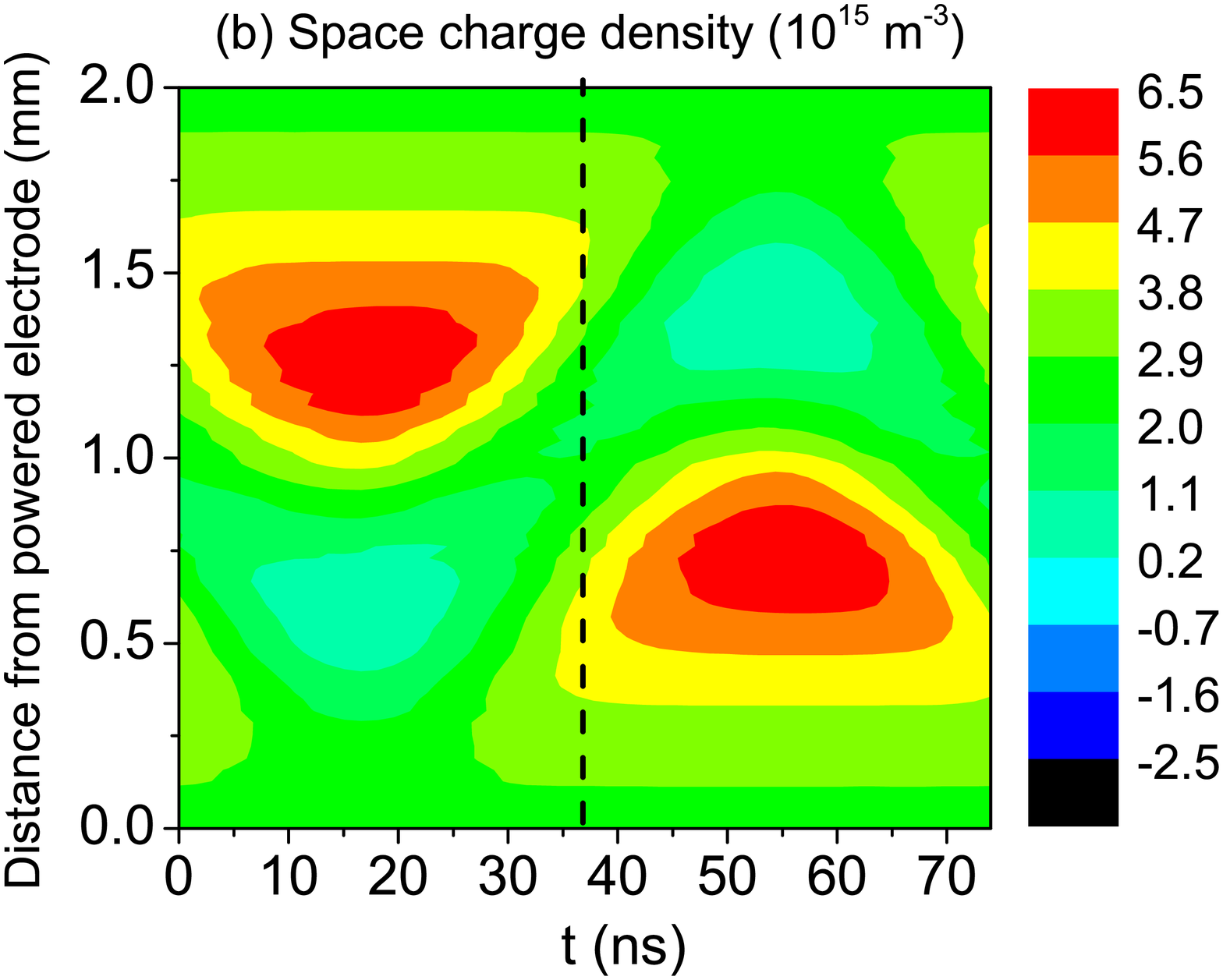}\\
  \includegraphics[width=0.39\textwidth]{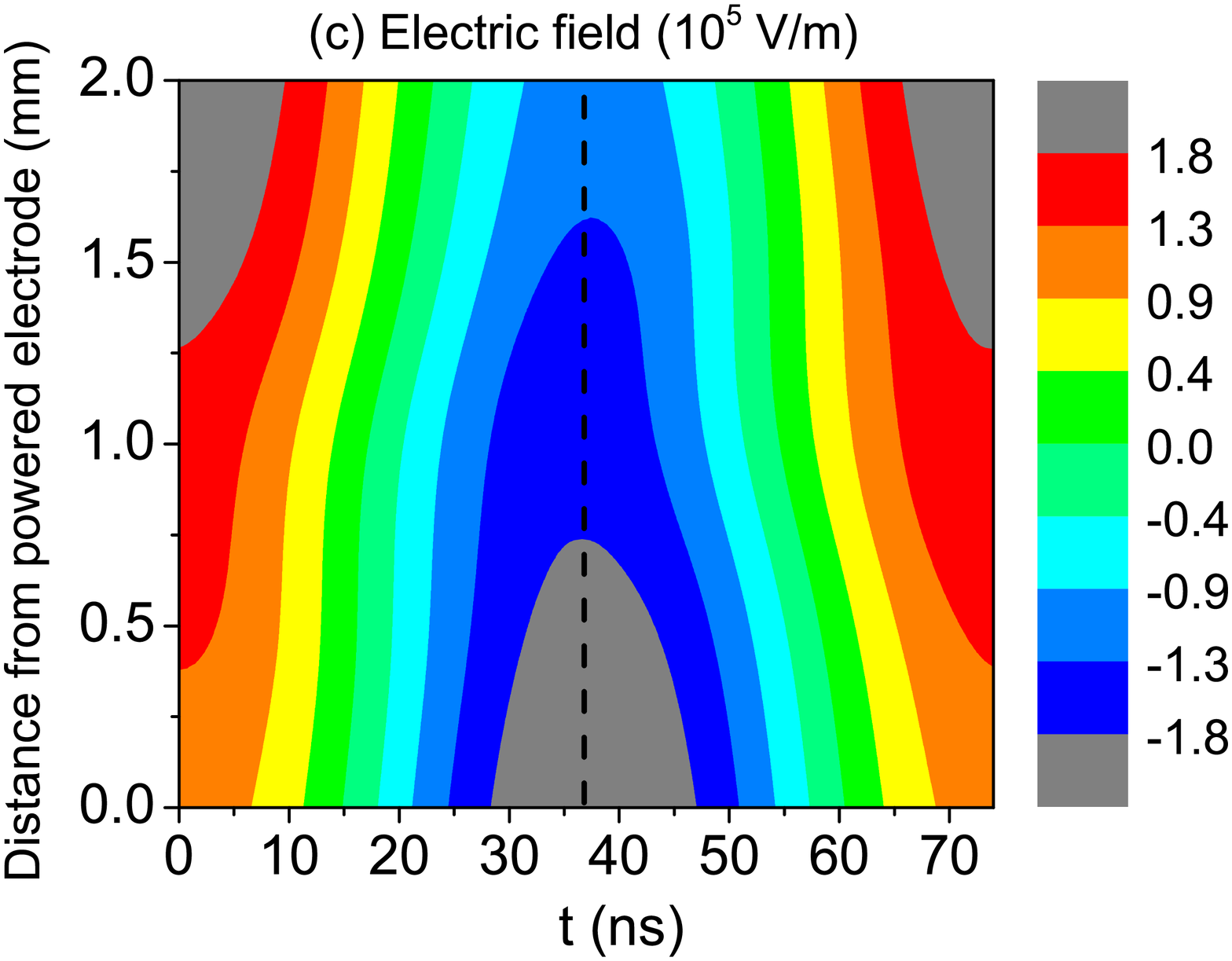}
&
   \includegraphics[width=0.39\textwidth]{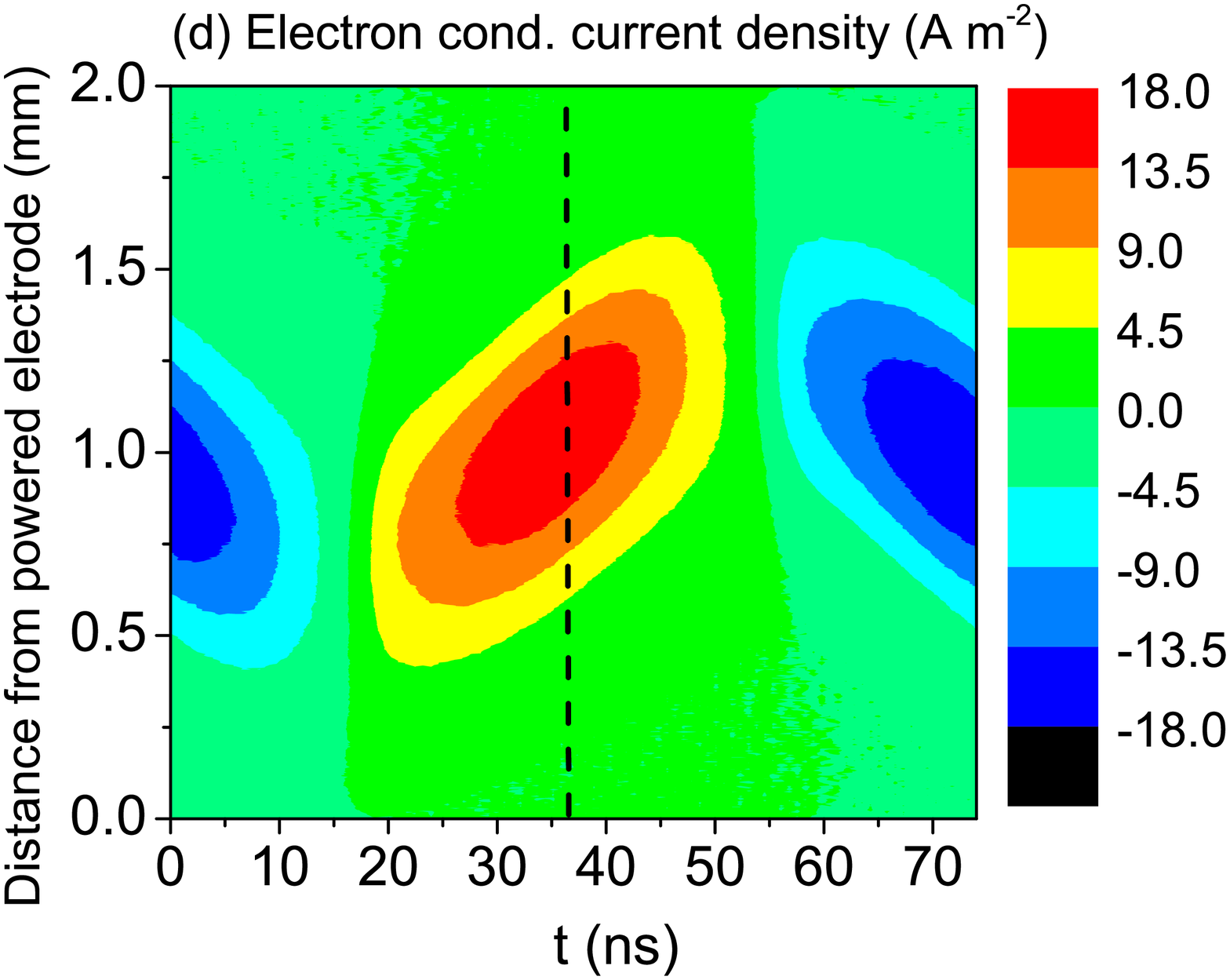}\\
   \includegraphics[width=0.39\textwidth]{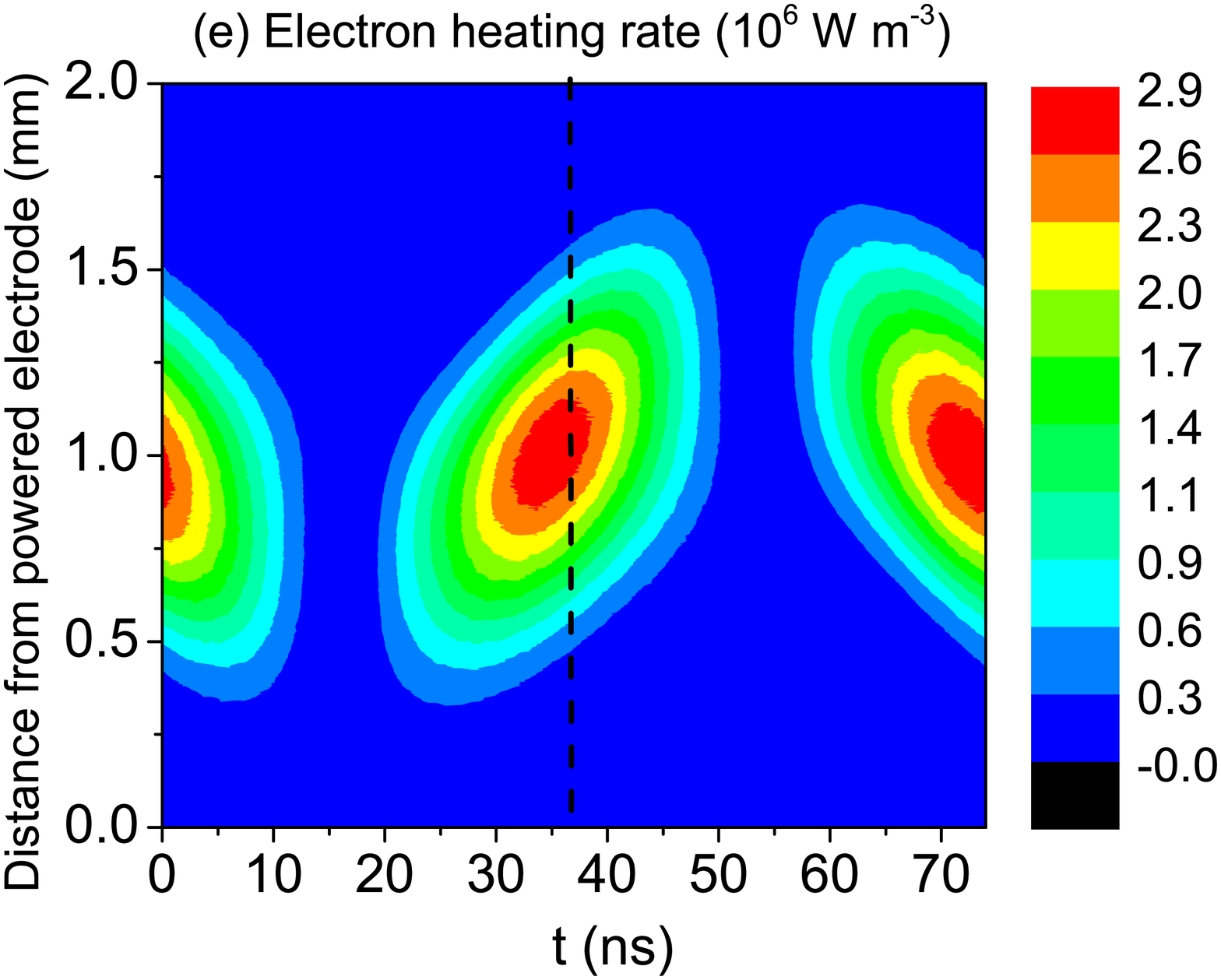}
&
   \includegraphics[width=0.39\textwidth]{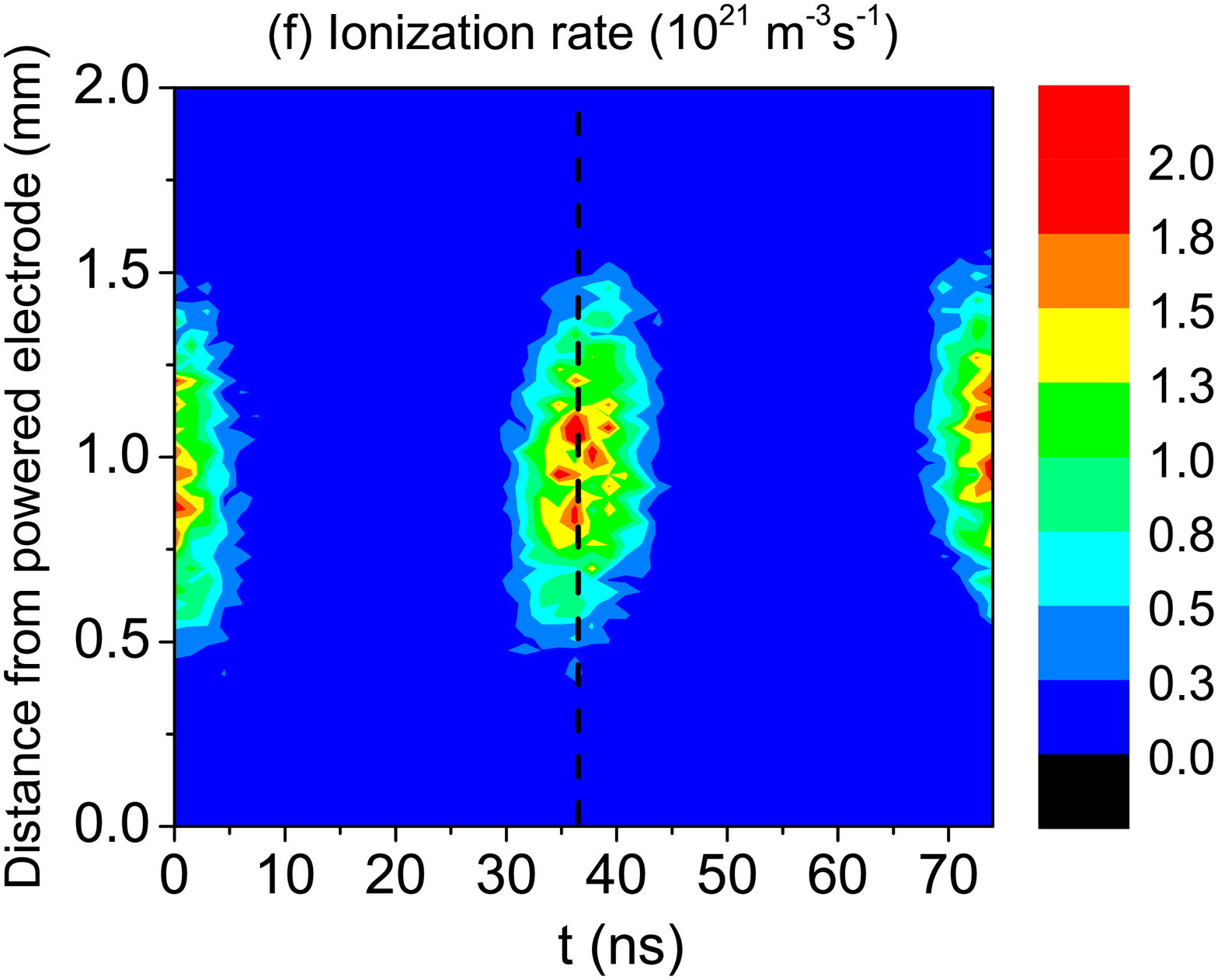}\\
      \includegraphics[width=0.39\textwidth]{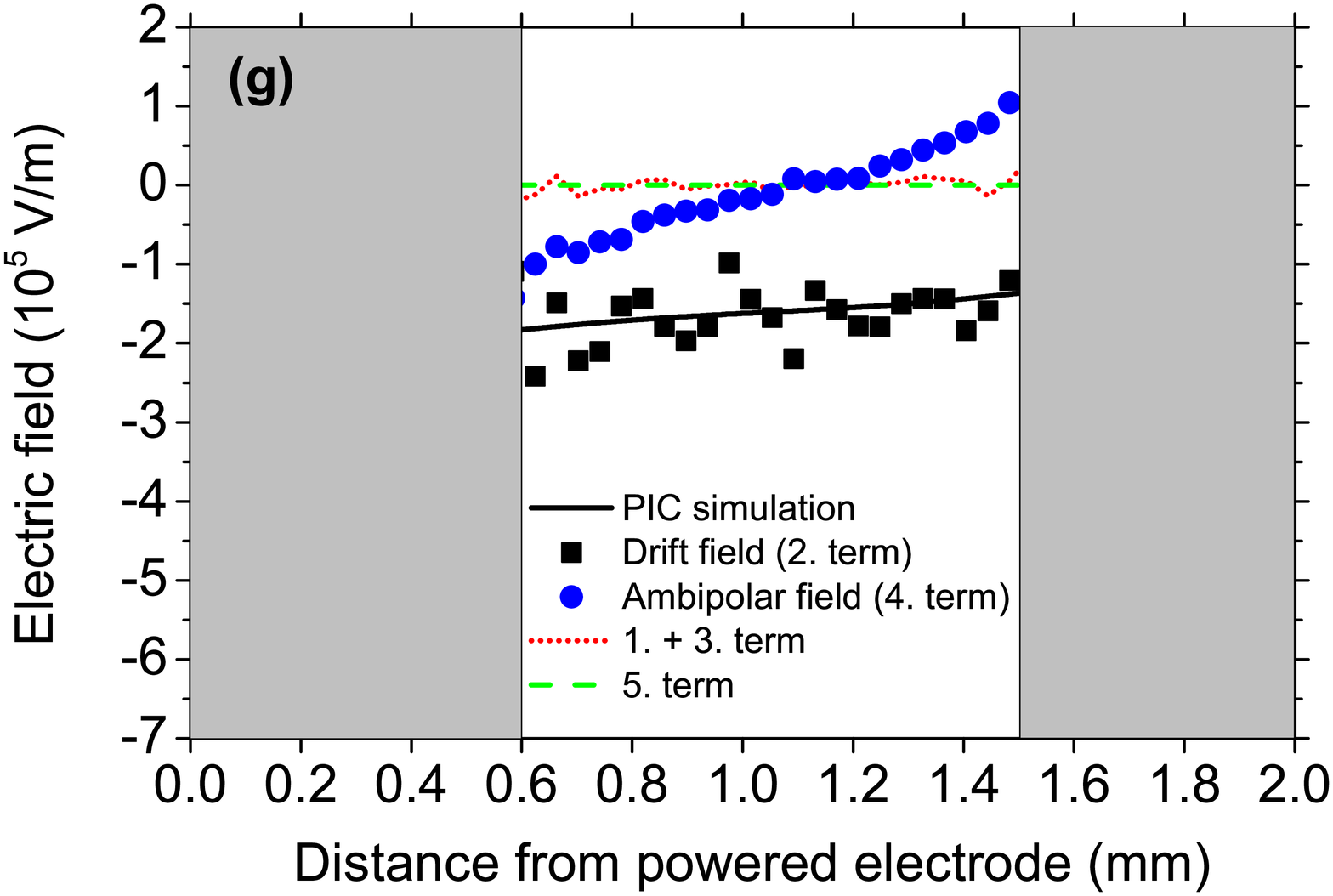}
&
   \includegraphics[width=0.39\textwidth]{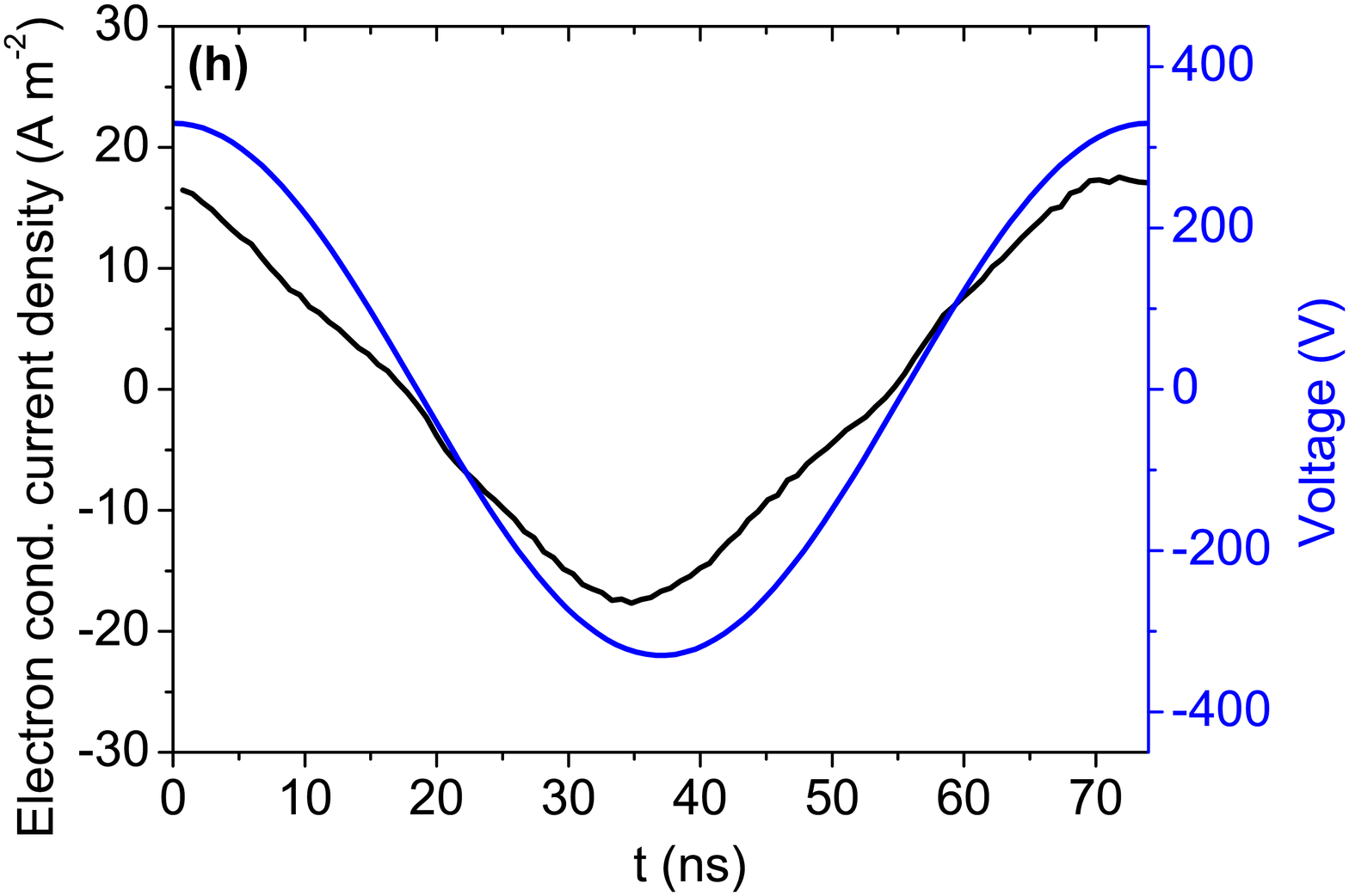}
\end{tabular}
\caption{PIC simulation results (He, 1 atm, 13.56 MHz, 2 mm gap, 330 V): Spatio-temporal plots of the (a) electron density, (b) space charge density, (c) electric field, (d) electron conduction current density, (e) electron heating, and (f) ionization rate. (g) shows the electric field profile at the time of max. ionization [vertical dashed lines in (a) - (f)] and the individual terms of eq. (\ref{EqEfield}). (h) shows the conduction current density in the discharge center and the applied voltage as a function of time.}
\label{Plots330V}
\end{center}
\end{figure}

Decreasing the driving voltage amplitude from 500 V to 330 V in case of atmospheric pressure microdischarges  operated in helium under otherwise identical discharge conditions as shown in figure \ref{Plots500V} significantly affects the electron heating and ionization dynamics, as shown in figure \ref{Plots330V}. The electron density is approximately one order of magnitude lower (a) and the electric field (c), the electron conduction current density (d), the electron heating rate (e), and the ionization rate (f) are maximum in the bulk at a later phase within the RF period. This is caused by a higher bulk resistance due to the lower plasma density, which now dominates compared to the sheath impedance and reduces the phase shift between current and voltage to $3.6^\circ$, i.e. the discharge becomes more resistive \cite{Phase1,Phase2}. Since current and voltage are almost in phase, maximum bulk electric field, the current, the heating, and the ionization are observed at a later phase within the RF period. The electron heating is again dominated by ohmic bulk heating. The heating and ionization rates peak in the discharge center and no longer at the sheath edges, since the conduction current is only high in the centre. The high electric field in the center is still predominantly caused by a low DC conductivity due to the high collision frequency, i.e. the second term of equation (\ref{EqEfield}). However, under these conditions there is a significant violation of quasineutrality, which increases towards the electrodes. As the applicability of the model is reduced in such regions, deviations between the modelled field and the electric field obtained from the simulation are found outside the center. 

\begin{figure*}[floatfix,h!]
\begin{center}
\includegraphics[width=0.75\textwidth]{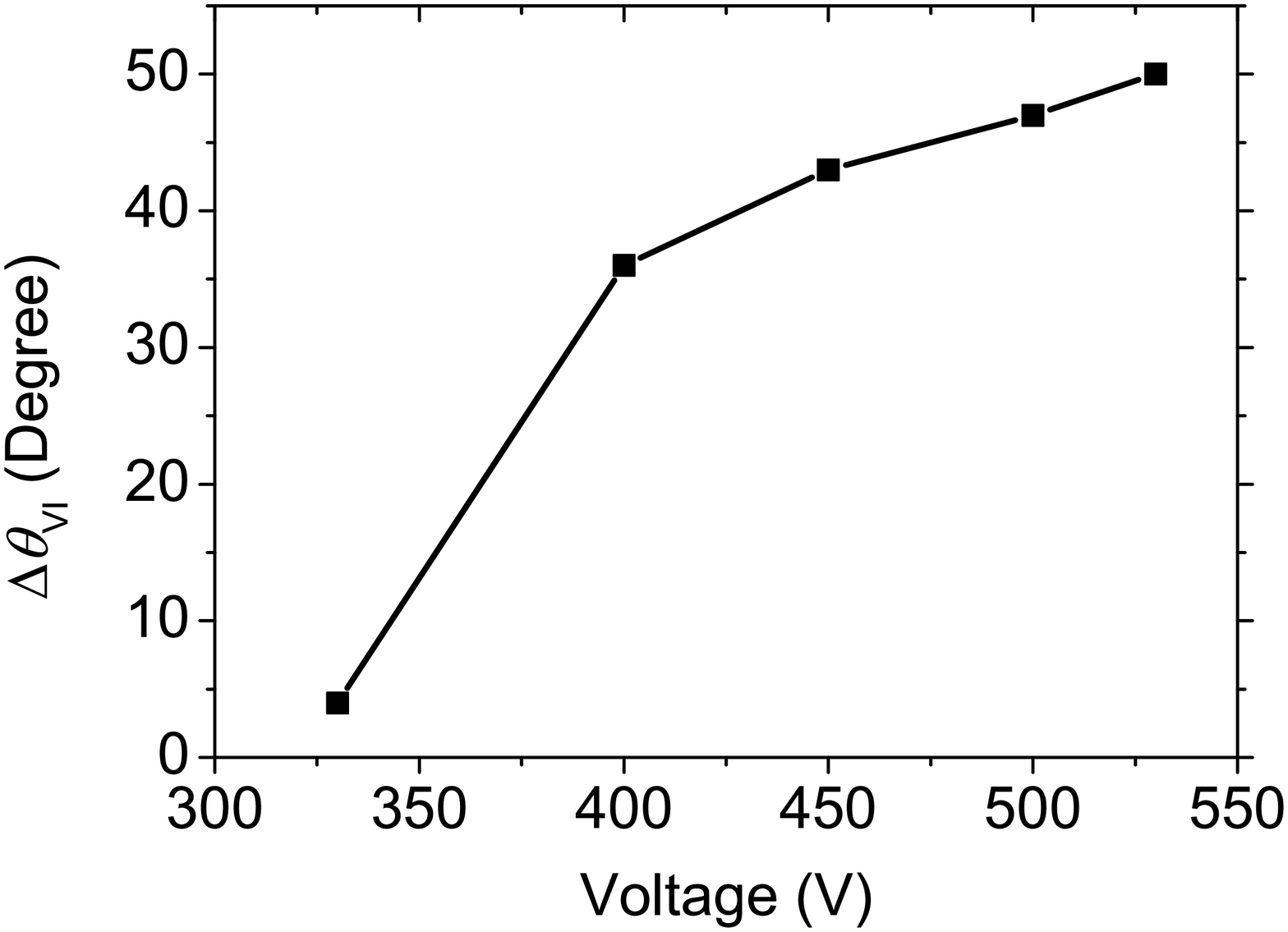}
\caption{\label{fig:5} Phase shift between the applied voltage 
and the conduction current density in the discharge center, $\Delta \theta_{\rm VI}$, as a function of the voltage amplitude (PIC simulation: He, 1 atm, 13.56 MHz, 2 mm gap).}
\label{PhaseShift}
\end{center}
\end{figure*}

Figure \ref{PhaseShift} shows the phase shift between the applied voltage 
and the conduction current density in the discharge center, $\Delta \theta_{\rm VI}$, as a function of the voltage amplitude between 330 V and 530 V. The phase shift is found to increase monotonically due to the increasing plasma density and decreasing resistivity. At voltages above 530 V the PIC simulation diverges. This might correspond to arcing in the experiment.

%%%%%%%%%%%%%%%%%%%%%%%%%%%%%%%%%%%%%%%%%%%%%%%%%
\newpage
\section{Conclusions}
Electron heating and ionization dynamics in capacitively coupled atmospheric pressure microplasmas operated in helium at 13.56 MHz and different voltage amplitudes were investigated by PIC simulations and semi-analytical modeling. The results were compared to electropositive argon and electronegative CF$_4$ macroscopic low pressure capacitive RF discharges. The electron heating dynamics in atmospheric pressure microplasmas is found to be dominated by the ohmic bulk heating of electrons due to high electric fields in the discharge center at the phases of maximum current in the RF period. The model reveals the physical origin of this $\Omega$-mode by identifying the high electric fields in the discharge center to be predominantly drift fields caused by a low conductivity due to the high electron-neutral collision frequency at atmospheric pressure. This heating mode is similar to the Drift-Ambipolar heating of electrons in low pressure electronegative macroscopic capacitive discharges operated in CF$_4$, where a high bulk electric field is caused by a low conductivity due to a low electron density caused by the high electronegativity. 

In atmospheric pressure microplasmas the heating and ionization dynamics are found to be affected by the amplitude of the driving voltage waveform. At high amplitudes, the plasma density is high and the bulk resistance is comparable to the sheath impedance, so that current and voltage are approximately 45$^\circ$ out of phase. Maximum ionization is observed adjacent to the sheath edges due to local maxima of the drift field caused by the ion density profile, that decreases towards the electrodes. These maxima of the ionization are not caused by direct interaction of electrons with the oscillating sheath electric fields such as observed in low pressure macroscopic electropositive discharges. There is no cooling of electrons during sheath collapse. At lower driving voltage amplitudes, the plasma density decreases, so that the bulk resistance increases and the discharge becomes more resistive. Consequently, the phase shift between voltage and current decreases and maximum ionization is observed in the discharge center at a later phase within the RF period.

These results might improve the understanding of the spatio-temporal emission in 
microscopic APPJs \cite{Schulz2007,Schaper2009,Benedikt2010,Kong2008} 
measured by Phase Resolved Optical Emission Spectroscopy 
\cite{PROES} and serve as a basis for a better understanding of chemical processes in such plasmas. Future investigations of the $\Omega$-mode in electronegative atmospheric pressure microplasmas as well as experimental verifications of the observed dependence of the phase shift between current and voltage on the driving voltage amplitude are clearly required.   

%\pagebreak
%%%%%%%%%%%%%%%%%%%%%%%%%%%%%%%%%%%%%%%%%%%%%%%%%
\ack
This project is funded by the DFG (German Research Foundation) within the framework of the Research Unit FOR 1123, SFB TR 87, SFB TR 24, project B5, and the Hungarian Fund for Scientific Research (OTKA K77653 and K105476).

%\pagebreak
%%%%%%%%%%%%%%%%%%%%%%%%%%%%%%%%%%%%%%%%%%%%%%%%%

%\clearpage

\end{document}